\documentclass[nohyper,12pt,letterpaper]{JHEP}  
\usepackage{epsfig}

\newfont{\frak}{eufm10 scaled 1200}

\newcommand\atmp[3]{{\it Adv. Theor. Math. Phys. }{\bf #1} (#2) #3}
\newfont{\Bbb}{msbm10 scaled 1200}     
\newcommand{\mathbb}[1]{\mbox{\Bbb #1}}
\DeclareSymbolFont{AMSa}{U}{msa}{m}{n}
\DeclareSymbolFont{AMSb}{U}{msb}{m}{n}
\let\Box\relax
\DeclareMathSymbol{\Box}{\mathord}{AMSa}{"03}

\def\IZ{{\mathbb Z}}
\def\IR{{\mathbb R}}

\def \eqn#1#2{\begin{equation}#2\label{#1}\end{equation}} 

\title{Note on the Quantum Mechanics of M Theory}

\author{Ofer Aharony and Tom Banks\\
  Department of Physics and Astronomy\\
  Rutgers University, Piscataway, NJ 08855-0849\\
E-mail: \email{oferah@physics.rutgers.edu, banks@physics.rutgers.edu}}

\abstract{ We observe that the existence of black holes limits the
extent to which M~Theory (or indeed any quantum theory of gravity) can
be described by conventional quantum mechanics.  Although there is no
contradiction with the fundamental properties of quantum mechanics,
one can prove that expectation values of Heisenberg operators at fixed
times cannot exist in an ordinary asymptotic Lorentz frame.  Only
operators whose matrix elements between the vacuum and energy
eigenstates with energy greater than the Planck scale are artificially
cut off, can have conventional Green's functions.  This implies a
Planck scale cutoff on the possible localization of measurements in
time. A similar behavior arises also in ``little string theories''. We
argue that conventional quantum mechanics in light cone time is
compatible with the properties of black holes if there are more than
four non-compact flat dimensions, and also with the properties of
``little string theories''. We contrast these observations with what
is known about M~Theory in asymptotically Anti-de Sitter spacetimes.
}


\received{???????? ?st, 1998}
\accepted{???????? ?th, 1998}

\preprint{\hepth{9812237}\\RU-98-52}
\begin{document}


\section{Introduction}

It is universally believed that M Theory is described by ordinary
quantum mechanics.   In this paper we will present evidence to the
contrary.  However, the modification of quantum mechanics we propose is
very mild and indeed the formalism we will use to investigate this
question is the standard formalism of the quantum theory.  The
only conventional axiom that we have to drop is the implicit one which allows
us to define Heisenberg operators, given an initial Hilbert space and
Hamiltonian.  We argue that in asymptotically 
Minkowski space time, this cannot be done
in M Theory (or any other candidate quantum theory of gravity) 
for any timelike asymptotic Killing vector.  This is a consequence of
the Bekenstein-Hawking formula for black hole entropy.  The same
considerations tell us why it {\it is} possible to construct a standard
Hamiltonian quantum mechanics for M Theory in asymptotically Anti-de
Sitter (AdS) spaces.  

We argue that when the number of Minkowski dimensions is greater than
four, Hamiltonian quantum M theory in the light cone frame is still
sensible.  We also note in passing that the success of light cone string
theory (in contrast with temporal gauge quantization of the string) is
related to the Hagedorn density of states of the string.

In precisely four noncompact dimensions, the light cone formalism
fails to cope with the density of states of black holes.  This
suggests that there may not even be a light cone Hamiltonian formalism
for nonperturbative M Theory in this framework.  Matrix Theory has
already provided evidence for the possible breakdown of the light cone
description of M Theory with four noncompact dimensions.  However, in
Matrix Theory this appeared to be related to the extra
compactification of a lightlike circle.  It is not clear to what
extent these problems are related to the (much milder) difficulties we
will point out here.  A light-cone description of the four dimensional
theory in terms of a non-standard quantum theory (such as a little
string theory) is not ruled out by our considerations; in the Matrix
Theory context such a description arises already when there are six
non-compact dimensions.

We also use similar methods to analyze non-gravitational little string
theories, and we conclude that they also do not have an ordinary
quantum mechanical description in the usual time frame. However, they
can (and do) have such a description in a light-cone frame.

\section{The Spectrum of a Hamiltonian, and Heisenberg Operators}

In a generally covariant theory, the definition of time and time
translation must be based on a physical object.  In noncompact spacetime
with appropriate asymptotic boundary conditions, appropriate physical
objects are the frozen classical values of the asymptotic spacetime
metric, and other fields.  We will restrict attention to such
asymptotically symmetric spaces in this paper.

The quantum theory lives in a Hilbert space which carries
a representation of the asymptotic
symmetry group, and among its generators we find the time translation
operator\footnote{For asymptotically Minkowski spaces we will ignore the
full Bondi-Metzner-Sachs symmetry group in this paper and imagine that
a particular
Poincare subgroup of it has been chosen by someone wiser than
ourselves. Perhaps one should think instead of the {\it Spi} group of
asymptotic diffeomorphisms of spacelike infinity, where a natural
Poincare subgroup exists.}.  
In the Minkowski case there is, up to conjugation, a
unique choice of time translation generator, while in the AdS case there
are two interesting inequivalent choices.  

Having chosen a Hamiltonian, we are now ready to discuss Heisenberg
operators, naively defined by 
\eqn{heisop}{O(t) \equiv e^{i H t} O e^{-i H t}.}
To test the meaning of this definition we compute the two point
(Wightman) function; the ground state expectation value of the product
of two operators at different times.  
By the usual spectral reasoning it has the form
\eqn{spectrep}{W(t) = <0\vert O^{\dagger}(0) O(t)\vert 0> =
\int_0^{\infty} dE e^{- i E t} \rho_O (E),}
where the spectral weight is defined by:
\eqn{spectwt}{\rho_O (E) = \sum_n \delta (E - E_n) {\vert 
<0\vert O \vert n >\vert}^2}
and is manifestly a sum of positive terms.  

The crucial issue is now the convergence of this formal Fourier
transform.  In interacting quantum field theory, an operator localized in a
volume $V$ will typically have matrix elements between the vacuum and
almost any state localized in $V$.  The only restrictions will come from
some finite set of global quantum numbers.  
At very high energies the density of states in volume $V$ is controlled
by the UV fixed point theory of which the full theory is a relevant
perturbation. Scale invariance and extensivity dictate that 
it has the form
\eqn{densst}{\rho (E) \sim e^{c V^{1\over d} E^{{d-1 \over d}}}}
for some constant $c$, where $d$ is the
spacetime dimension.  As a consequence, even if an operator $O$ has
matrix elements between the vacuum and states of arbitrarily high energy
(and indeed, even if these matrix elements grow as the exponential of a
power of the energy less than one, which is not typical), the Fourier
integral converges and defines a distribution\footnote{This may be
seen by analytic continuation to Euclidean space.}.  

In local field theory there are special operators whose matrix elements
between the vacuum and high energy states are highly suppressed.  These
are operators which are linear functionals of local fields of fixed
dimension at the UV fixed point.  The spectral function of such fields
has power law dependence on the energy (as opposed to the exponential
dependence implied by (\ref{densst})), which means that most of their
matrix elements to most high energy states vanish rapidly with the energy.
A typical operator localized in volume
$V$ is a nonlinear functional of such fields.  Using the operator
product expansion we can write a formal expansion of 
it in terms of the linear fields, but it
will generically contain large contributions from fields of arbitrarily
large dimension, and its spectral function will have the asymptotics of
the full density of states.  

Physically, we can understand the special role of local operators by
first thinking of a massive field theory at relatively low energy.
The intermediate states of energy $E$ created by a local field will be
outgoing scattering states of a number of massive particles bounded by
$E/m$, whose momenta point back to the place where the field
acts. When $E$ is large these will be very special states among all
those available in the same volume and with the same energy.  A
localized burst of energy will dissipate rather than thermalize.  As
$E$ gets very large, this description becomes inadequate. However,
note that if we choose a basis of local fields of fixed dimension in
the UV conformal field theory, then any given field in the basis
creates only states in a single irreducible representation of the
conformal algebra.  The degeneracy of these representations is much
smaller than that of the full theory.

Now we want to ask the same question for M Theory compactified to $d$
asymptotically Minkowski spacetime dimensions.  As in our discussion of
field theory, we work in a fixed asymptotic Lorentz frame and discuss
the time evolution operator appropriate to that frame.  We claim that
the high energy density of states in superselection sectors with finite
values of all Lorentz scalar charges, is dominated by $d$ dimensional
Schwarzschild black holes.  The density of such black hole states is, by
the Bekenstein-Hawking formula,
\eqn{bhd}{\rho_{BH} \sim e^{c (E/M_P)^{{(d-2) \over (d-3)}}},}
where $M_P$ is the $d$ dimensional Planck mass and $c$ is a constant.
The Fourier integral no longer converges unless the matrix elements of
the operator vanish rapidly for large $E$ \footnote{or unless the
operator connects the vacuum only to a very small fraction of the black
hole states.  Given the thermal nature of the black hole ensemble, {\it
i.e.} the fact that it consists primarily of states with the same gross
properties, operators of the latter type do not correspond to very
realistic probes of the system.}.  Note furthermore that if we employ an
operator with an energy cutoff much larger than the Planck mass (the
point above which the behavior (\ref{bhd}) presumably sets in), 
then for all times longer than the inverse
cutoff, the integral will be completely dominated by the states at the
cutoff energy scale.  Thus the only kinds of operators whose Green
functions exhibit the usual property that the variation over a time
scale $\Delta T$ probes excitations of energy $1/ \Delta T$ are those
with an energy cutoff below $M_P$.  

The necessity of cutting off operators implies a non-locality in the
physics of M Theory at the time scale $M_P^{-1}$.  Indeed, a probe of
the system localized at time $t = t_0$ differs from one localized at
$t=0$ because it couples to the operator $\int dE e^{ i E t_0} O(E)$
rather than the same integral with $t_0$ set to zero.  If $t_0 \ll
M_P^{-1}$ and $O(E)$ is cut off at $E \sim M_P$ then these operators are
indistinguishable. 

An important feature of the density of states (\ref{bhd}), which
distinguishes it from that of field theory, is its independence of the
volume. This is related to the familiar instability of the thermal
ensemble in quantum gravity to the formation of black holes \cite{gpy}.  In
ordinary field theory, the typical state of very high energy $E$ on a
torus is a member of the translation-invariant thermal ensemble.
However, as a consequence of the Jeans instability, any attempt to
create a translationally invariant state with finite energy density in a
theory containing gravity will fail.  Once the size of a patch of finite
energy density exceeds its Schwarzschild radius, it collapses into a
black hole.  So, the generic high energy state in
M~Theory with four or more  asymptotically flat dimensions is a single
black hole.  

This gives us some insight into the question raised in the previous
footnote of why there are no analogs in M~Theory of local operators of fixed
dimension which couple only to a few of the high energy states.  In
quantum gravity an operator carrying a very large 
energy $E$ cannot create a state
more localized than the corresponding Schwarzschild radius.  The most
localized states of a given energy are also the generic states of that
energy and they are black holes.  

Some readers familiar with the AdS/CFT correspondence \cite{juan} will
at this point be objecting that these considerations seem to
contradict the successful description of M Theory in AdS spaces by
quantum field theories.  Other AdS/CFT {\it cognoscenti} will have
already recognized that in fact that correspondence is another example
of the rules we have just promulgated.  AdS spaces have two
inequivalent types of interesting time evolutions, the Poincare and
the global time.  The appropriate black objects which dominate the
high energy density of states for these two definitions of energy are
near extremal black branes, and AdS Schwarzschild black holes
respectively.  Both of these have positive specific heat, that is the
density of states is an exponential of a power of the energy less than
one.  This is completely consistent with quantum field theory, and of
course the matching of the thermodynamics of these objects with that
of conformal field theories \cite{klebtsey} is one of the primary
clues which led to the AdS/CFT correspondence.

Note that the major discrepancy between the behavior of the density of
states in AdS and Minkowski spacetimes suggests that the extraction of
flat space physics from that of AdS may be quite subtle.  In
\cite{bdhmetc} it was suggested that in appropriate regions of parameter
space, the Hilbert space of the quantum field theories describing AdS
physics contain an energy regime describing flat space black holes whose
Schwarzschild radius is much smaller than the radius of curvature of AdS.
Somehow one must find observables in the AdS theory which probe only
this energy regime and reduce to the corresponding flat space S-matrix.

To summarize: we have argued that in asymptotically Minkowski space
time, M Theory cannot be described as a conventional quantum theory.
Although the violation of the rules of quantum mechanics appears mild,
we would like to emphasize that any quantum system which corresponds
to the quantization of a classical action which is the integral over
time of a Lagrangian by Euclidean path integrals would appear to have
well defined Green functions.  Thus, although the systems which might
describe M Theory in Minkowski space violate the abstract rules of
quantum mechanics only by having a bizarre spectral density for the
Hamiltonian, they are unlikely to have a conventional Lagrangian
description with respect to an ordinary time variable.
  
\section{Light Cone Time and Black Holes}

String theory has traditionally been formulated in light cone gauge
because, for various technical reasons, one was unable to find another
gauge in which the system  was obviously a unitary quantum theory.
More recently, the light cone gauge has been seen \cite{thorn} as the
framework in which the holographic \cite{thooftsuss} nature of string
theory and M Theory becomes apparent.  Matrix Theory \cite{bfss} is a
nonperturbative formulation of M Theory in Discrete Light Cone
Quantization (DLCQ) on spaces with at least $6$ 
Minkowski dimensions.   

In this section we would like to propose that the success of Hamiltonian
quantization in light cone gauge is partly due to the absence in light
cone gauge of the problems described in the previous section.  This
leads us to anticipate a problem with any light cone 
formulation of the theory in 4 noncompact dimensions.

The argument is extremely simple.  Light cone energy is defined by
\eqn{lcen}{{\cal E} = { {\bf P}^2 + M^2 \over P^+}.}
For fixed longitudinal momentum and vanishing transverse momentum, we
can therefore write the density of black hole states in light cone energy
as 
\eqn{lcden}{\rho ({\cal E}) \sim e^{ ({P^+ {\cal E} \over M_P^2})^{{d-2
\over 2(d - 3)}}}.}
Thus, for $d > 4$ the density of states is well behaved, and we can
hope to describe the system as some sort of conventional quantum
mechanics.  
In precisely four dimensions the Bekenstein-Hawking formula implies a
Hagedorn spectrum for M Theory in light cone energy.  

It is extremely
interesting to compare these observations with the problems encountered
in compactified DLCQ M Theory, or Matrix Theory.
There it is known that for 7 or more noncompact dimensions the DLCQ
description is a quantum field theory, while in 6 dimensions it is
\cite{brs,lst} a
{\it little string theory}.  
As described in the next section,
the little string theory has a Hagedorn spectrum.   
Finally, for five dimensions the theory
involves gravity, at least in the simplest
maximally SUSY case \cite{seibetal}. 
Note that, qualitatively, the DLCQ
density of states seems to mirror that of the uncompactified light cone
theory with two fewer dimensions.  It is field theoretical down to $d=7$
and has a Hagedorn form for $d=6$.  
There does not, however, appear to be any quantitative mapping of one
problem on to the other.  The exponents in the energy-entropy relation
are completely different.  Indeed, it is well known \cite{tbrev} that
the DLCQ theory contains many states which must decouple in the large
$N$ limit if the limiting theory is to be Lorentz invariant.  These
undoubtedly are responsible for the dramatically different behavior of
the density of states in the DLCQ and uncompactified light cone
descriptions.  The Lorentz invariant theory contains only the states
with energy of order $1/N$ in the large $N$ limit, and the asymptotic
density of states in this theory refers to the asymptotics of the {\it
coefficient} $1/N$ in the DLCQ theory.  
  
With five noncompact dimensions, the lightcone Hamiltonian of toroidally
compactified $DLCQ_N$ M~Theory is that of 11D SUGRA in the presence of an
$A_N$ singularity. In this picture the time is that of the rest frame of
the singularity.  The theory is compactified on a six torus with radii
of order the Planck scale.  For finite $N$, the high energy density of
states of this theory is dominated by 5D black holes and a conventional
Heisenberg quantum mechanics will not exist.  

On the other hand, it has been suggested by Kachru, Lawrence, and
Silverstein (KLS) \cite{kls} that DLCQ
M~Theory compactified on a Calabi-Yau manifold will have fewer and more
innocuous extraneous states.  These authors also propose that the Matrix
description of this theory might be a $3+1$ dimensional field theory.
Note that the entropy of such a field theory scales like $E^{3/4}$, {\it
which is precisely the scaling with light cone energy of the entropy of
five dimensional black holes.}  In the present authors' opinion, this
observation supports the suggestion of \cite{kls} and
should encourage us to search for the relevant $3+1$ dimensional field
theory.   One clue to its nature might be the fact that the field
theoretical nature of the spectrum seems to survive the large $N$ limit.
In maximally SUSY Yang Mills theory, there are degrees of freedom with
energies as low as $N^{-1/3}$ and a $3+1$ dimensional spectrum, but
there do not appear to be any field theoretical degrees of freedom with
energies as low as $1/N$.  The KLS field theory should have such states.

It is well known that with four noncompact dimensions ({\it i.e.} two
transverse dimensions) the DLCQ theory ceases to exist.  The light cone
Hamiltonian is the rest frame Hamiltonian
 of toroidally wrapped D7 branes in weakly coupled
IIB string theory.  This is only a picturesque description, for it is
not self-consistent.  If we assume an asymptotically locally flat space,
the D7 branes have infinite energy (by a BPS formula).  However, the
gravitational back reaction is infinite and one merely learns that the
asymptotically flat ansatz is not self-consistent.  Quite likely the
theory does not exist at all.  The DLCQ theory is really a
compactification to $2+1$ flat dimensions.  Furthermore, like all light
cone theories it describes only excited states of the vacuum rather than
the vacuum itself.  If a 2+1 dimensional theory
has four or more supercharges (so that there is an exactly massless
scalar in the SUGRA multiplet) then we do not expect there to be many
such states.  A generic ``localized'' excitation of the vacuum
creates a geometry which is not asymptotically flat.  Thus the Hilbert
space of the system with asymptotically flat boundary conditions (or
even asymptotically locally flat) is very small and contains only a few
topological excitations \cite{bs2d}.   

This problem, which arises in DLCQ,
does not appear to have much to do with the apparent absence
of a Heisenberg quantum mechanics in the lightcone theory compactified to
4 dimensions.  The problem there is an excess of states which do satisfy
the asymptotic boundary conditions, rather than a lack of such states.
Thus while Matrix Theory may be a useful guide to many properties of
M~Theory, we cannot expect to get the physics of low dimensional
compactifications right without finding a light-cone description of
the Lorentz invariant theory (without compactifying a light-like circle).  
  
\subsection{A Loose End}

A possible objection to the above discussion is that black holes are not
stable. Thus they do not really correspond to the eigenspectrum of the
Hamiltonian. However, the lifetime of black holes goes to infinity as a
power of their mass.  They are thus extremely narrow resonances and
correspond to an enhancement of the density of scattering states.
In principle we could have formulated our description of the
behavior of Green's functions in terms of complete sets of scattering
states and the properties of the S-matrix.

\section{Hamiltonian Description of Little String Theories}

In section 2 we briefly discussed the convergence of the formal
expression (\ref{spectrep}) in the case of local field theories, and
noted that there appeared to be no problem with it in this
case. However, it was realized in the past few years that there are
also non-local field theories which can be decoupled from gravity, in
particular little string theories. Although decoupled from
gravity, these behave in many ways like critical string theories. In this
section we will analyze the behavior of these theories at high
energies, and we will argue that they have a Hagedorn density of
states. Therefore, using the arguments of section 2, it is not clear
how to define local operators in these theories using the usual time
variable, but a light-cone description of these theories in terms
of an ordinary Lagrangian quantum mechanics does seem to make sense. We
will discuss in detail only the case of little string theories
with 16 supercharges in 6 dimensions of type $A_{k-1}$, but we expect
the conclusions to be more general.

The construction of little string theories with 16 supercharges in
6 dimensions was discussed in \cite{lst}. The original
definition of these theories involved looking at $k$ NS 5-branes in
type IIA (for the ${\cal N}=(2,0)$ little string theories) or type
IIB (for the ${\cal N}=(1,1)$ little string theories) string
theories, and taking the limit of $g_s \to 0$ with the string scale
$\alpha'$ constant. While this construction provides evidence for the
existence of such little string theories, 
it does not allow for explicit computations in these
theories. Two independent methods for making direct computations in
these theories were developed in the past two years, and we will use
both of them to compute the density of states at
high energies.

\subsection{The Equation of State from the Holographic Description}

The first method we will use is the holographic description of the
little string theories \cite{abks}, which is a generalization of
the AdS/CFT correspondence \cite{juan,gkp,wittenads}. The little string
theories are claimed to be dual to a background of M theory or
string theory which, at a large radial coordinate, asymptotes to a
linear dilaton background of string theory (with a string metric which
is the standard metric on $\IR^7\times S^3$). It is easy to generalize
this description also to the case of finite energy density or
temperature. As in the case of the AdS/CFT correspondence
\cite{wittentemp}, the relevant background (at least at high energy
densities) is the near-horizon background of near-extremal
NS5-branes. The simplest way to derive this background is just to take
the $g_s \to 0$ limit in the background of a near-extremal 5-brane
\cite{malstro}. The result is very similar to the background described
in \cite{abks}, but with the linear dilaton direction and the time
direction replaced by an $SL(2)/U(1)$ black hole (with an appropriate
$SL(2)$ level so that the total central charge is ${\hat c}=10$). 
The black hole
background also has a varying dilaton with the string coupling going
to zero far from the horizon. If we start with an energy density $E/V
= \mu M_s^6$ on the 5-brane, the string coupling at the horizon is
$g_s^2 \sim k/\mu$. Since for large $k$ the curvature of this background
is small, it follows that for $\mu \gg k \gg 1$ we can trust the
supergravity description of this background
\cite{malstro}, and it provides a holographic dual for the
little string theories with this energy density\footnote{For smaller
energy densities we expect this background to be corrected, at least
in the ${\cal N}=(2,0)$ case where at low temperatures and 
energy densities
the solution should become localized in the eleventh direction
\cite{imsy}.}. 

In particular, we can use this description to compute the 
equation of state of
the little string theories from the Bekenstein-Hawking formula (as
was done for the case of the AdS/CFT correspondence in 
\cite{wittentemp}), and we
find that the equation of state is \cite{malstro}
\eqn{lststate}{E = {M_s \over \sqrt{6k}} S,}
where $E$ and $S$ can be taken to be either the total energy and
entropy or the energy and entropy densities (since the formula does
not depend on the volume). As was noted in a similar context in
\cite{juanold}, this is the same density of states as that of a free
string theory with a string scale of $k\alpha'$ and $c=6$, or of a
free string theory with a string scale of $\alpha'$ and $c=6k$, even
though the theory does not seem to be a free string theory; an
interpretation as a theory with a string scale $\alpha'$ seems more
likely since the theory has a T-duality symmetry corresponding to this
scale \cite{lst}. In any case, for our purposes it is enough to note
that at high energy densities we get the equation of state
(\ref{lststate}), which is a Hagedorn density of states with a Hagedorn
temperature of $T_H = M_s / \sqrt{6k}$ (signifying that the canonical
ensemble can only be defined below this temperature). 

The fact that
(at least for large volumes compared to the string scale) 
the equation of state does not depend on
the volume suggests 
that, as in M~theory, 
the high
energy density of states is dominated by the states of a single
object.  In little string theory we believe that the analog of the black
hole is a single long string. This seems to be 
the message of the Hagedorn spectrum.
Note that this is not strictly
true in free string theory, since there
the numbers of strings in each string
state are an infinite set of conserved quantities. However,
when interactions are turned on, multiple string states can convert into
a single long string and this has more entropy.  In the interacting
little string theory we should expect this phenomenon to occur as well.
We will provide some supporting evidence for this from the DLCQ
description below, by computing the Hagedorn description via an
independent argument which shows that it can naturally arise just from
single string states.
As noted above, the 
corresponding phenomenon for black holes in asymptotically Minkowski
spacetime is the Gross Perry Yaffe \cite{gpy} instability of the
translation invariant thermal ensemble to the formation of a
single large black hole.
In the little string theories we do not expect a similar localization
of the generic high energy states, but they still seem to correspond
to single objects, unlike the local field theory case.

We argued for the full M~Theory that the existence of black holes
precluded the existence of local operators, which couple only to a small
subset of the high energy states.  We believe that the same is
true in the little string theory, as a consequence of the fact that the
generic high energy state is a single big string. 
In field theory (on a large but finite torus) the generic state of high
energy is a translation-invariant gas.   But in an interacting string
gas in any finite volume, once the energy is taken large enough, the
density of strings is such that overlaps are inevitable, and in the
presence of interactions the high entropy single string state will be
preferred.   In a gas, it is easy to construct operators which create
only one of the constituents from the vacuum.  On the other hand, if
perturbative string theory is any guide, it is very difficult to
construct operators with matrix elements between the vacuum and only a
few of the highly degenerate excited string states.  

Our arguments here have necessarily been quite heuristic because we do
not have a good description of the eigenstates of little string
theories. Nonetheless, combined with the supporting evidence from the
DLCQ and holographic pictures, and especially the calculation of Peet
and Polchinski \cite{pp} which suggests that correlation
functions of little string theory operators in momentum space do not
have Fourier transforms, our description of the physics of this system
seems plausible.

\subsection{The Equation of State from the DLCQ Description}

A completely different description of the little string theories
is their discrete light-cone quantization, which was described in
\cite{abkss,wittenlst} for the case with ${\cal N}=(2,0)$
supersymmetry and in \cite{ganset} for the case with ${\cal
N}=(1,1)$. In both cases the description of the theory with light-like
momentum $P^+=N/R$ is given by a $1+1$ dimensional conformal theory
with $c=6Nk$, compactified on a circle of radius $\Sigma = 1 / R
M_s^2$. Conformal invariance dictates the equation of state of these
theories at high energies (above the scale $1/\Sigma$) to be 
\eqn{eqstateone}{E_{DLCQ} = 6 N k \Sigma T^2; \qquad S = 6 N k \Sigma T,}
or
\eqn{eqstatetwo}{E_{DLCQ} = S^2 / 6 N k \Sigma.}
As in the previous section, we can easily translate this into an
equation of state for the full space-time theory (a similar procedure
was carried out for Matrix black holes in \cite{mbh}). In the absence of
transverse momentum we have $E_{DLCQ} P^+ = E^2$, so we get
\eqn{eqstatethree}{E = \sqrt{E_{DLCQ} N / R} = {M_s \over \sqrt{6k}}
S,}
which is exactly the same relation as (\ref{lststate}). Note that all
factors of $N$ and $R$ dropped out of this expression, as well as any
dependence on the volume of space, without the necessity of taking the
large $N$ limit; this happens here because of
special properties of the DLCQ of the little string theories and
would not be true in general.
 
The computation above gives the high-energy density of states in the
DLCQ theory; unfortunately this is not what we are interested in for
the Lorentz-invariant theory, for which only states whose energies are
of order $1/N$ are relevant (obviously for large $N$ these energies
will become smaller than the scale $1/\Sigma$ above which our previous
computation was valid). Luckily, as in the case of the DLCQ
description of type IIA string theory \cite{motl,bansei,dvv}, 
we can argue that these
theories have states whose energy scales like $1/N$ which obey the
same equation of state as the full theory. In the case of type IIA
string theory, these states involved ``long string'' states which
changed by a $U(N)$ gauge transformation when going around the compact
circle. For free type IIA string theory the DLCQ involved a free CFT
of central charge ${\hat c}=8N$, while the ``long string states'' (for the
lowest-lying states for which the gauge transformation was equivalent
to a permutation of order $N$ of the eigenvalues of the $U(N)$ adjoint
matrices) involved a CFT of central charge ${\hat c}=8$ but compactified on
a circle of radius ${\tilde \Sigma} = N \Sigma$; since the formula
(\ref{eqstatetwo}) depends only on the product $c\Sigma$ these states
obey the same equation of state as that of the full theory. We would
like to suggest that a similar mechanism holds also in the DLCQ
description of the little string theories. For the ${\cal
N}=(2,0)$ case this involves the Higgs branch of a $U(N)$ gauge theory
while for the ${\cal N}=(1,1)$ case it involves the Coulomb branch of
a $U(N)^k$ gauge theory. At least in the latter case it is clear that
the theory includes ``long string'' states with energies of order
$1/N$ just like in the full type IIA string theory, and it seems
likely that the central charge for the theory describing the ``long
strings'' will be $1/N$ of the total central charge. A complete
analysis of these ``long string'' states will be presented
elsewhere. Unfortunately, since these states are strongly interacting
(unlike the case discussed above of weakly coupled
type IIA string theory), it is not clear if we can
really trust this description for computing the density of
states. In particular, it is not obvious that the ``long string''
states are adequately described by a local CFT.
However, it seems plausible that ``long string'' 
states do exist and obey an
equation of state similar to (\ref{eqstatetwo}) (up to possible
numerical factors). We view this as additional evidence for the
validity of (\ref{lststate}) in the little string theories, and for
the entropy being dominated by single-string states.

\subsection{Discussion}

Let us now discuss the consequences of (\ref{lststate}) for the
description of the little string theories. As discussed in section
2, this behavior implies that correlation functions of standard
Heisenberg operators do not exist in these theories, at least when the
time difference between the operators is smaller than the Hagedorn
scale $1/T_H$.  Indeed, using the holographic description \cite{abks} of
the little string theories, Peet and Polchinski \cite{pp} have provided
independent evidence that the correlation functions of these theories
are not Fourier transformable and do not obey the rules of quantum field
theory.  This supplements the arguments based on T-duality.

Thus, we expect the DLCQ description of these theories (or perhaps a
direct large $N$ limiting version of it) to be the only Lagrangian
quantum theory which computes the correlation functions and
eigenspectrum of the theory.  

We want to emphasize the way in which the DLCQ analysis agrees with the
Bekenstein-Hawking analysis of these theories.  DLCQ predicts a Hagedorn
spectrum in ordinary Lorentz frames in a very robust way.  The argument
depends only on general properties of $1+1$ dimensional field
theories. The only possible loophole in the argument is that the
spectrum of states whose energies in the large $N$ limit are of order
$1/N$ 
might not be field theoretic.  We believe we have provided plausible
arguments which close this loophole, though more work is necessary to
elucidate the nature of long string states in these interacting
theories.  The success of the Bekenstein-Hawking argument in predicting
the correct density of states in these systems motivated us to apply it
to quantum gravity in the bulk of this paper.

\section{Conclusions and Questions}

What are we to make of the failure of Heisenberg quantum mechanics in
light cone gauge for gravitational theories
in four dimensions?  Does this spell the end of the
search for a nonperturbative Lagrangian formulation of M~Theory ?
There are several possibilities:
\begin{itemize}
\item 1. Our reaction to the nonexistence of Green's functions has been
too violent, at least in the case of a Hagedorn spectrum.  For example,
in first quantized string theory, the system has a Hagedorn spectrum in
ordinary Lorentz frames.  Nonetheless a covariant Lagrangian and
Hamiltonian formalism exists in conformal gauge.  The time variable in
this formalism is not connected to {\it any} spacetime variable.  There
are two reasons to be suspicious of the possibility of generalizing this
formalism to a nonperturbative interacting theory.  First of all, 
the light cone gauge
formulation of the theory does not have any problems, and the covariant
formalism bears a very close resemblance to it.  Secondly, the free
string theory is
completely integrable and there are natural operators which communicate
only with single states among the Hagedorn spectrum.  
One should however emphasize as well that the divergence of Green's
functions is much less dramatic for a Hagedorn spectrum than it is for
black holes far from extremality.  In particular, in the Hagedorn case,
Euclidean Green's functions exist as long as all time intervals are
sufficiently long.  Furthermore, the little string theories give us an
example of interacting, non-gravitational systems with a Hagedorn
spectrum. 

\item 2. Perhaps, as in Matrix Theory, light cone M~Theory in four
dimensions can be formulated as some sort of compactification of a
theory with an auxiliary Lorentz invariance under which the light cone
time variable of M~Theory transforms as the time component of a
Lorentz vector.  Then we can formulate the theory in the light cone
frame of the auxiliary Lorentz group and deal with the Hagedorn
spectrum by the same trick which works in first quantized string
theory.  This is the proposal of \cite{abkss,wittenlst} for treating
little string theories.  It is not clear how such a proposal could
work for four dimensional M theory without compactifying a light-like
direction (which we cannot do in this case as discussed in section
3). In a DLCQ theory, the auxiliary Lorentz group relates the
lightcone Hamiltonian to the charges of longitudinally wrapped
branes. These extra unwanted ``momentum'' quantum numbers disappear
into the ultraviolet in the large $N$ limit, so the auxiliary Lorentz
group does not act on the states which survive in the large $N$ limit
(the ``momentum'' states all have an infinite energy in the limit of a
non-compact light-cone description). Thus, this symmetry 
should not exist in the exact
light-cone description of the four dimensional theory.
Nonetheless, we should emphasize again that
the closest analog to a putative four dimensional light cone M~Theory is
the little string theory, and this {\it does} have a DLCQ description
which is an ordinary quantum mechanical theory.
 
\item 3. Perhaps M~Theory with four Minkowski dimensions can only be
defined as the limit of M~Theory with AdS asymptotics.  We have 
pointed out above that the current understanding of the AdS/CFT
correspondence does not furnish us with a prescription for extracting
the Minkowski S-Matrix from the AdS theory, but perhaps this difficulty
can be overcome. The most serious objection we can find to such a proposal 
is that the most likely candidate theory would be of the form $AdS_2
\times S^2 \times X$, but $AdS_2$ theories seem to be topological
\cite{strom}.

\item 4. The real world is not Minkowski space but rather a cosmological
space time.  Perhaps we should be searching for the fundamental
formulation of M~Theory only in the context of closed cosmologies
(cosmologies where all space-like slices are compact). In
this case we do not expect the notion of Hamiltonian to have a
fundamental significance.  Time evolution is a concept which is recovered
only in a semiclassical approximation.  The problems we seem to have
with formulating a Heisenberg picture quantum mechanics may signal a
breakdown of this semiclassical approximation rather than a fundamental
problem.   We have to admit that we don't understand how this could be
the case.
\end{itemize}

Whatever the resolution of these difficulties, we cannot end this paper
without making note of two significant points. 
The first is the privileged position of four dimensions in
this discussion.  Gravitational theories with fewer than four
(Minkowski) dimensions do not have many states.  This has been advanced
in \cite{bs2d} as a reason why they are not the endpoint of cosmological
evolution.  On the other hand, the Bekenstein-Hawking formula tells us
that in some sense four dimensional Minkowski spacetimes have more
states at a given asymptotic 
energy level than higher dimensional spaces (and all
Minkowski spaces have more states than any AdS space).  Perhaps this
observation will be the key to understanding why the world we observe is
four dimensional.

Our final comment is to emphasize the similarity between the high energy
spectra of four dimensional light cone M~Theory and of little string
theories in an ordinary reference frame.  This suggests that, although
the light cone quantum mechanics describing four dimensional M~Theory
is not a conventional Lagrangian
theory, it may be some sort of little string theory.
This fascinating conjecture is an obvious direction
for future work.

\acknowledgments
We are grateful to M. Berkooz, W. Fischler, D. Kutasov, N. Seiberg
and Sasha Zamolodchikov for valuable discussions. This work 
was supported in part by the DOE under grant
number DE-FG02-96ER40559. 



\begin{thebibliography}{19}        

%
\bibitem{gpy} D.\, J.\, Gross, M.\, J.\, Perry, L.\, G.\, Yaffe,
{``Instability of Flat Space at Finite Temperature,''}
\prd{25}{1982}{330}.
\bibitem{juan} J.\, M.\, Maldacena,
{``The Large $N$ Limit of Superconformal Field Theories and
Supergravity,''}
\atmp{2}{1998}{231}, \hepth{9711200}.
\bibitem{klebtsey} S.\, S.\, Gubser, I.\, R.\, Klebanov, A.\, W.\,
Peet,
{``Entropy and Temperature of Black 3-branes,''}
\prd{54}{1996}{3915}, \hepth{9602135};
I.\, R.\, Klebanov, A.\, A.\, Tseytlin,
{``Entropy of Near Extremal Black $p$-branes,''}
\npb{475}{1996}{164}, \hepth{9604089}. 
\bibitem{bdhmetc} T.\,Banks, M.\, Douglas, G.\, Horowitz, E.\, Martinec,
{``AdS Dynamics from Conformal Field Theory,''}
\hepth{9808016}.
\bibitem{thorn} C.\, Thorn,
{``Reformulating String Theory with the $1/N$ Expansion,''} 
published in Sakharov Conf. on Physics, Moscow (1991) 447,
\hepth{9405069}. 
\bibitem{thooftsuss} G.\, 't Hooft, 
{``Dimensional Reduction in Quantum Gravity,''}
published in Salamfest 1993, 284,
\grqc{9310026};
L.\, Susskind, {``The World as a Hologram,''}
\jmp{36}{1995}{6377},
\hepth{9409089}.
\bibitem{bfss} T.\, Banks, W.\, Fischler, S.\, Shenker, L.\, Susskind, 
{``M~Theory as a Matrix Model: A Conjecture,''}
\prd{55}{1997}{5112}, 
\hepth{9610043}.
\bibitem{brs} M.\, Berkooz, M.\, Rozali, N.\, Seiberg,
{``Matrix Description of M Theory on $T^4$ and $T^5$,''}
\plb{408}{1997}{105}, \hepth{9704089}.
\bibitem{lst} N.\, Seiberg,
{``New Theories in Six Dimensions and Matrix Description of M Theory
on $T^5$ and $T^5/\IZ_2$,''}
\plb{408}{1997}{98}, \hepth{9705221}.
\bibitem{seibetal} N.\, Seiberg,
{``Why is the Matrix Model Correct ?''}
\prl{79}{1997}{3577},
\hepth{9710009}.
\bibitem{tbrev} T.\, Banks, {``Matrix Theory,''} 
\npps{67}{1998}{180},
\hepth{9710231}.
%
\bibitem{kls} S.\, Kachru, A.\, Lawrence, E.\, Silverstein, 
{``On the Matrix Description of Calabi-Yau Compactifications,''} 
\prl{80}{1998}{2996}, \hepth{9712223}.
\bibitem{bs2d} T.\, Banks, L.\, Susskind,
{``The Number of States of Two Dimensional Critical String Theory,''} 
\prd{54}{1996}{1677}, \hepth{9511193}.
\bibitem{abks} O.\, Aharony, M.\, Berkooz, D.\, Kutasov, N.\, Seiberg,
{``Linear Dilatons, NS5-branes and Holography,''}
\jhep{9810}{1998}{004}, \hepth{9808149}.
\bibitem{gkp} S.\, S.\, Gubser, I.\, R.\, Klebanov, A.\, M.\,
Polyakov,
{``Gauge Correlators from Noncritical String Theory,''}
\plb{428}{1998}{105}, \hepth{9802109}.
\bibitem{wittenads} E.\, Witten,
{``Anti-de Sitter Space and Holography,''}
\atmp{2}{1998}{253}, \hepth{9802150}.
\bibitem{wittentemp} E.\, Witten,
{``Anti-de Sitter Space, Thermal Phase Transition, and Confinement in
Gauge Theories,''}
\atmp{2}{1998}{505}, \hepth{9803131}.
\bibitem{malstro} J.\, M.\, Maldacena, A.\, Strominger,
{``Semiclassical Decay of Near Extremal Five-branes,''}
\jhep{9712}{1997}{008}, \hepth{9710014}.
\bibitem{imsy} N.\, Itzhaki, J.\, M.\, Maldacena, J.\, Sonnenschein,
S.\, Yankielowicz,
{``Supergravity and the Large $N$ Limit of Theories with Sixteen
Supercharges,''}
\prd{58}{1998}{46}, \hepth{9802042}.
\bibitem{juanold} J.\, M.\, Maldacena,
{``Statistical Entropy of Near Extremal Five-branes,''}
\npb{477}{1996}{168}, \hepth{9605016}.
\bibitem{pp} A.\, W.\, Peet, J.\, Polchinski,
{``UV/IR Relations in AdS Dynamics,''}
\hepth{9809022}.
\bibitem{abkss} O.\, Aharony, M.\, Berkooz, S.\, Kachru, N.\, Seiberg,
E.\, Silverstein,
{``Matrix Description of Interacting Theories in Six Dimensions,''}
\atmp{1}{1998}{148}, \hepth{9707079}.
\bibitem{wittenlst} E.\, Witten,
{``On the Conformal Field Theory of the Higgs Branch,''}
\jhep{9707}{1997}{003}, \hepth{9707093}.
\bibitem{ganset} O.\, J.\, Ganor, S.\, Sethi,
{``New Perspectives on Yang-Mills Theories with Sixteen
Supersymmetries,''}
\jhep{9801}{1998}{007}, \hepth{9712071}.
\bibitem{mbh} T.\, Banks, W.\, Fischler, I.\, R.\, Klebanov, L.\,
Susskind,
{``Schwarzschild Black Holes from Matrix Theory,''}
\prl{80}{1998}{226}, \hepth{9709091}; \jhep{9801}{1998}{008},
\hepth{9711005}.
\bibitem{motl} L.\, Motl,
{``Proposals on Nonperturbative Superstring Interactions,''}
\hepth{9701025}.
\bibitem{bansei} T.\, Banks, N.\, Seiberg,
{``Strings from Matrices,''} \npb{497}{1997}{41}, \hepth{9702187}.
\bibitem{dvv} R.\, Dijkgraaf, E.\, Verlinde, H.\, Verlinde,
{``Matrix String Theory,''} \npb{500}{1997}{43}, \hepth{9703030}.
\bibitem{strom} A.\, Strominger, {``$AdS_2$ Quantum Gravity and String
Theory,''} \hepth{9809027}; J.\, Maldacena, J.\, Michelson, A.\,
Strominger, {``Anti-de Sitter Fragmentation,''} \hepth{9812073}. 
%
\end{thebibliography}
\end{document}